\documentclass[twoside,twocolumn,9pt]{article}
\usepackage[utf8]{inputenc}
\usepackage{extsizes}
\usepackage[super,sort&compress,comma]{natbib} 
\usepackage[version=3]{mhchem}
\usepackage[left=1.5cm, right=1.5cm, top=1.785cm, bottom=2.0cm]{geometry}
\usepackage{balance}
\usepackage{times,mathptmx}
\usepackage{sectsty}
\usepackage{graphicx} 
\usepackage{lastpage}
\usepackage[format=plain,justification=justified,singlelinecheck=false,font={stretch=1.125,small,sf},labelfont=bf,labelsep=space]{caption}
\usepackage{float}
\usepackage{fancyhdr}
\usepackage{fnpos}
\usepackage[english]{babel}
\addto{\captionsenglish}{%
  
}
\usepackage{amsmath}
\usepackage{amssymb}
\DeclareMathOperator{\Tr}{Tr}
\usepackage{array}
\usepackage{droidsans}
\usepackage{charter}
\usepackage[T1]{fontenc}
\usepackage[usenames,dvipsnames]{xcolor}
\usepackage{setspace}
\usepackage[compact]{titlesec}
\usepackage{hyperref}
\usepackage{epstopdf}

\definecolor{cream}{RGB}{222,217,201}

\begin{document}

\pagestyle{fancy}
\thispagestyle{plain}
\fancypagestyle{plain}{

\renewcommand{\headrulewidth}{0pt}
}

\makeFNbottom
\makeatletter
\renewcommand\LARGE{\@setfontsize\LARGE{15pt}{17}}
\renewcommand\Large{\@setfontsize\Large{12pt}{14}}
\renewcommand\large{\@setfontsize\large{10pt}{12}}
\renewcommand\footnotesize{\@setfontsize\footnotesize{7pt}{10}}
\makeatother

\renewcommand{\thefootnote}{\fnsymbol{footnote}}
\renewcommand\footnoterule{\vspace*{1pt}%
\color{cream}\hrule width 3.5in height 0.4pt \color{black}\vspace*{5pt}} 
\setcounter{secnumdepth}{5}

\makeatletter 
\renewcommand\@biblabel[1]{#1}            
\renewcommand\@makefntext[1]%
{\noindent\makebox[0pt][r]{\@thefnmark\,}#1}
\makeatother 
\renewcommand{\figurename}{\small{Fig.}~}
\sectionfont{\sffamily\Large}
\subsectionfont{\normalsize}
\subsubsectionfont{\bf}
\setstretch{1.125} 
\setlength{\skip\footins}{0.8cm}
\setlength{\footnotesep}{0.25cm}
\setlength{\jot}{10pt}
\titlespacing*{\section}{0pt}{4pt}{4pt}
\titlespacing*{\subsection}{0pt}{15pt}{1pt}

\fancyfoot{}
\fancyfoot[RO]{\footnotesize{\sffamily{1--\pageref{LastPage} ~\textbar  \hspace{2pt}\thepage}}}
\fancyfoot[LE]{\footnotesize{\sffamily{\thepage~\textbar\hspace{3.45cm} 1--\pageref{LastPage}}}}
\fancyhead{}
\renewcommand{\headrulewidth}{0pt} 
\renewcommand{\headrulewidth}{0pt}
\setlength{\arrayrulewidth}{1pt}
\setlength{\columnsep}{6.5mm}
\setlength\bibsep{1pt}

\makeatletter 
\newlength{\figrulesep} 
\setlength{\figrulesep}{0.5\textfloatsep} 

\newcommand{\topfigrule}{\vspace*{-1pt}%
\noindent{\color{cream}\rule[-\figrulesep]{\columnwidth}{1.5pt}} }

\newcommand{\botfigrule}{\vspace*{-2pt}%
\noindent{\color{cream}\rule[\figrulesep]{\columnwidth}{1.5pt}} }

\newcommand{\dblfigrule}{\vspace*{-1pt}%
\noindent{\color{cream}\rule[-\figrulesep]{\textwidth}{1.5pt}} }

\makeatother

\twocolumn[
  \begin{@twocolumnfalse}
\vspace{3cm}
\sffamily
\begin{tabular}{m{4.5cm} p{13.5cm} }

 &\noindent\LARGE{\textbf{Active nematic-isotropic interfaces in channels}} \\
\vspace{0.3cm} & \vspace{0.3cm} \\

 & \noindent\large{Rodrigo C. V. Coelho,$^{\ast}$\textit{$^{a,b}$} Nuno A. M. Araújo,\textit{$^{a,b}$} and Margarida M. Telo da Gama\textit{$^{a,b}$}} \\\\

&\noindent\normalsize{
We use numerical simulations to investigate the hydrodynamic behavior of the interface between nematic (N) and isotropic (I) phases of a confined active liquid crystal. At low activities, a stable interface with constant shape and velocity is observed separating the two phases. For nematics in homeotropic channels, the velocity of the interface at the NI transition increases from zero (i) linearly with the activity for contractile systems and (ii) quadratically for extensile ones. Interestingly, the nematic phase expands for contractile systems while it contracts for extensile ones, as a result of the active forces at the interface. Since both  activity and temperature affect the stability of the nematic, for active nematics in the stable regime the temperature can be tuned to observe static interfaces, providing an operational definition for the coexistence of active nematic and isotropic phases. At higher activities, beyond the stable regime, an interfacial instability is observed for extensile nematics. In this regime defects are nucleated at the interface and move away from it. The dynamics of these defects is regular and persists asymptotically for a finite range of activities. We used an improved hybrid model of finite differences and lattice Boltzmann method with multi-relaxation-time collision operator, the accuracy of which allowed us to characterize the dynamics of the distinct interfacial regimes.
}

\\

\end{tabular}

 \end{@twocolumnfalse} \vspace{0.6cm}

  ]

\renewcommand*\rmdefault{bch}\normalfont\upshape
\rmfamily
\section*{}
\vspace{-1cm}


\footnotetext{\textit{$^{a}$~Centro de Física Teórica e Computacional, Faculdade de Ciências,
Universidade de Lisboa, P-1749-016 Lisboa, Portugal; E-mail: rcvcoelho@fc.ul.pt}}
\footnotetext{\textit{$^{b}$~Departamento de Física, Faculdade de Ciências,
Universidade de Lisboa, P-1749-016 Lisboa, Portugal. }}





\section{Introduction}

Systems formed by units or particles capable of transforming the energy of the environment into directed motion are known as active matter. These intrinsically non-equilibrium systems are characterized by complex dynamical behavior. Examples that occur naturally are dense suspensions of bacteria, mixtures of microtubule-kinesin, and shoals of fish (see \cite{Ramaswamy_2017,BechingerRMP2016} and references therein). While a detailed description of a particular system is daunting, particle based simulations of simple models exploded in the last decade and reported a range of new phenomena. Motility induced phase separation (MIPS) is one of the most striking~\cite{CatesAnnRev2015}.

A different approach focused on the phenomenology observed in dense systems, which results from the interplay of self-propulsion and alignment interactions due to the shape of the particles, and their hydrodynamic coupling to the suspending medium~\cite{Doostmohammadi2018,Ramaswamy_2017,BechingerRMP2016,MarchettiRMP2013}. Here, we follow this route and consider momentum conserving (or wet~\cite{Doostmohammadi2016}) active systems with nematic order, which may be described in the hydrodynamic limit by the continuum equations of liquid crystals (LC) with an extra stress term that accounts for the 
activity. 
While passive LCs~\cite{p1995physics, beris1994thermodynamics} have been studied for decades, active nematics are currently a hot topic of 
fundamental research. Indeed, only recently, a second active stress term was proposed, which may have a profound effect on the behaviour and stability of active nematics~\cite{MaitraPNAS2018}.

One question that has been investigated concerns the conditions to observe directed coherent flow, which is key to applications 
in micromotors and micromachines~\cite{Wueaal1979, Opathalage4788}. While in the bulk active turbulence may be unavoidable for any degree of activity~\cite{Doostmohammadi2018,Ramaswamy_2017,BechingerRMP2016,MarchettiRMP2013}, confinement in narrow channels results in a spontaneous symmetry breaking and opens a window to observe directed flow at low but non-zero activities~\cite{VoituriezEPL2005, marenduzzo07steady, EdwardsEPL2009}. 
As the activity increases, the directed flow changes to oscillatory, dancing disclinations, and ultimately turbulent regimes, both in the deep nematic phase~\cite{ShendrukSM2017, Giomi_2012, Wensink14308} and also at temperatures above the passive nematic-isotropic transition~\cite{ChandragiriSoftMatter2019}.

The fact that the activity, by itself, induces local nematic ordering and even active turbulence, raises the question of the coexistence between active nematic and isotropic phases and of the behaviour of the interfaces between them. Although this is a topic of intense research for MIPS in active systems with scalar order
parameters~\cite{CatesAnnRev2015, SolonPRE2018, SolonNJP2018, TjhungPRX2018}, it has not been addressed for active nematics, despite the fact that an intrinsic free energy has been proposed for these systems~\cite{ThampiEPL2015}.

In addition to fundamental issues related to the existence and characterizaton of phase coexistence in active nematics,
interfaces of LCs have played a prominent role in applications. For example, they can be controlled easily in order to modify their
optical properties~\cite{Blow_2013} and provide an efficient means to transport colloids~\cite{refId0}. Interfaces of active LCs are expected to play similar roles. Previous studies of active and passive LCs have focused on the
dynamical behaviour of droplets of one in the other, with the observation that passive droplets may be driven through the active fluid while active droplets may rotate in passive nematics  ~\cite{PhysRevLett.113.248303, C4SM00937A, C7SM00325K, C7SM01019B}. The coexistence of
active nematic and isotropic phases and the dynamics of their interfaces have not been investigated. 
\begin{figure}[t]
\center
\includegraphics[width=\linewidth]{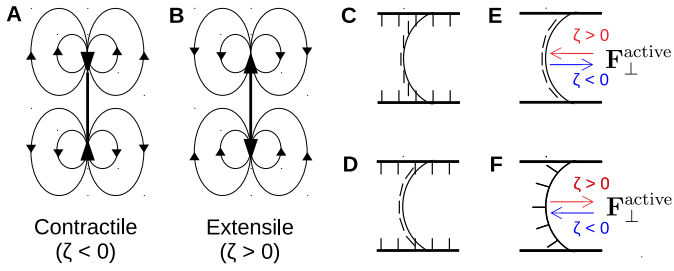}
\caption{A and B) Velocity field in a domain of a contractile and an extensile system. C) Directors in a narrow channel, where the strong anchoring at the walls dominates the planar interfacial anchoring. D) Directors in a channel, where the planar interfacial anchoring dominates. E) Direction of the active force, at the centre of the interface, for extensile and contractile systems with planar interfacial anchoring. F) Direction of the active force, at the centre of the interface, for extensile and contractile systems with perpendicular interfacial anchoring.  }
\label{active-particle-fig}
\end{figure}
\begin{figure}[b]
\center
\includegraphics[width=\linewidth]{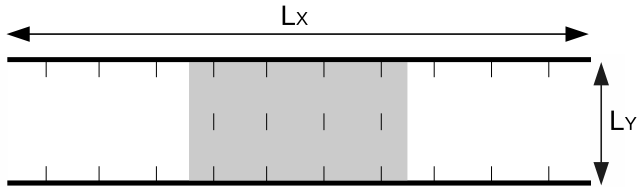}
\caption{Homeotropic channel setup and initial conditions. The nematic phase, with uniform director perpendicular to the walls, is set in the center of the channel with isotropic fluid elsewhere. The anchoring at the walls is fixed (strong) and perpendicular (homeotropic). Periodic boundary conditions are applied in the $x$ direction and no-slip at the walls. }
\label{setup-fig}
\end{figure}
In this paper, we consider the dynamics of the interface between the nematic and the isotropic phases of an active LC
confined in channels with strong homeotropic anchoring at the walls. The system exhibits a nematic ordering transition, driven by the temperature at fixed (low) activity. 
The velocity of the interface at the passive NI temperature increases from zero (i) linearly with the activity for contractile systems and (ii) quadratically for extensile ones. In the former, the nematic order increases while it decreases in the latter. This behavior is unexpected as extensile systems favor nematic order, but it is readily understood as a result of the active forces at the interface. The effect is reversed in planar channels as the active forces at the interface are also reversed, as we show here. At fixed activity, the velocity of the interface depends linearly on the temperature. In the stable regime the temperature can be tuned to observe a static interface in the channel providing an operational definition for the coexistence of confined active nematic and isotropic phases. Beyond the stable regime, extensile nematics exhibit an interfacial instability in hometropic and planar channels.

We performed computer simulations of the hydrodynamic equations of active liquid crystals using a hybrid method of
lattice Boltzmann~\cite{succi2018lattice, kruger2016lattice} (LB) and a finite differences (FD) predictor corrector~\cite{vesely2001computational}. 
There are challenges associated with the simulation of interfaces in multiphase LB models, such as the control of spurious velocities and the difficulty to set the physical values of both viscosity and surface tension~\cite{0953-8984-25-24-245103}. To overcome these problems, we introduced various improvements, as discussed in the following. 

Our paper is organized as follows. In Sec.~\ref{methods-sec}, we describe the equations of motion and the numerical method. In Sec.~\ref{stable-int-ec}, we describe the results of numerical simulations of passive and active nematic interfaces in channels and show how the active interfaces may become static through a shift in the temperature. In Sec.~\ref{flow-states-sec}, we discuss the interfacial instability observed in extensile nematics and mention briely the flow states that occur at higher activities. In Sec.~\ref{other-setups-sec}, we investigate the effects of anchoring at the channel walls and check the effect of the interaction between interfaces. In Sec.~\ref{conclusions} we summarize and conclude$^\dag$\footnotetext{\dag~Electronic Supplementary Information (ESI) available: details of the MRT model and videos of the interfacial dancing state and interfacial instability are given in the ESI. See DOI: 00.0000/00000000.}. 
\begin{figure}[t]
\center
\includegraphics[width=\linewidth]{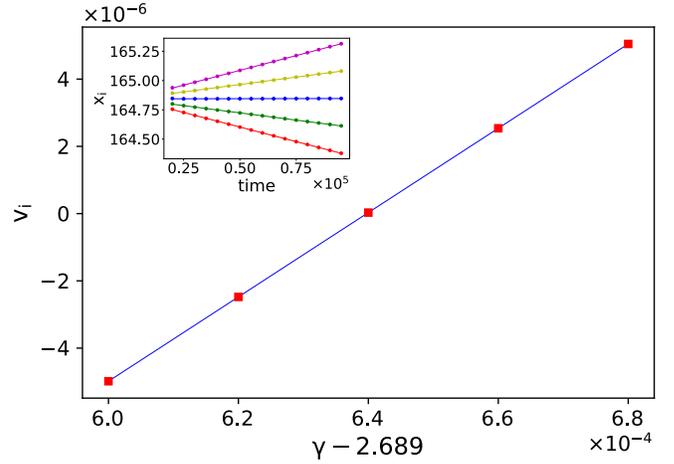}
\caption{Velocity of the (right) passive interface, $v_i$, at temperatures $\gamma$ close to $\gamma_{NI}=2.7$. The squares are the measured velocities and the solid line is a linear fit, which yields the coexisting temperature $\gamma_{NI}^{ch}$ where the passive interface in the channel is static. The inset shows the (right) interface position as a function of time at different values of $\gamma$ (from top to bottom): 2.68968, 2.68966, 2.68964, 2.68962, 2.68960.}
\label{passive-fig}
\end{figure}
\begin{figure*}[htb]
\center
\includegraphics[width=\linewidth]{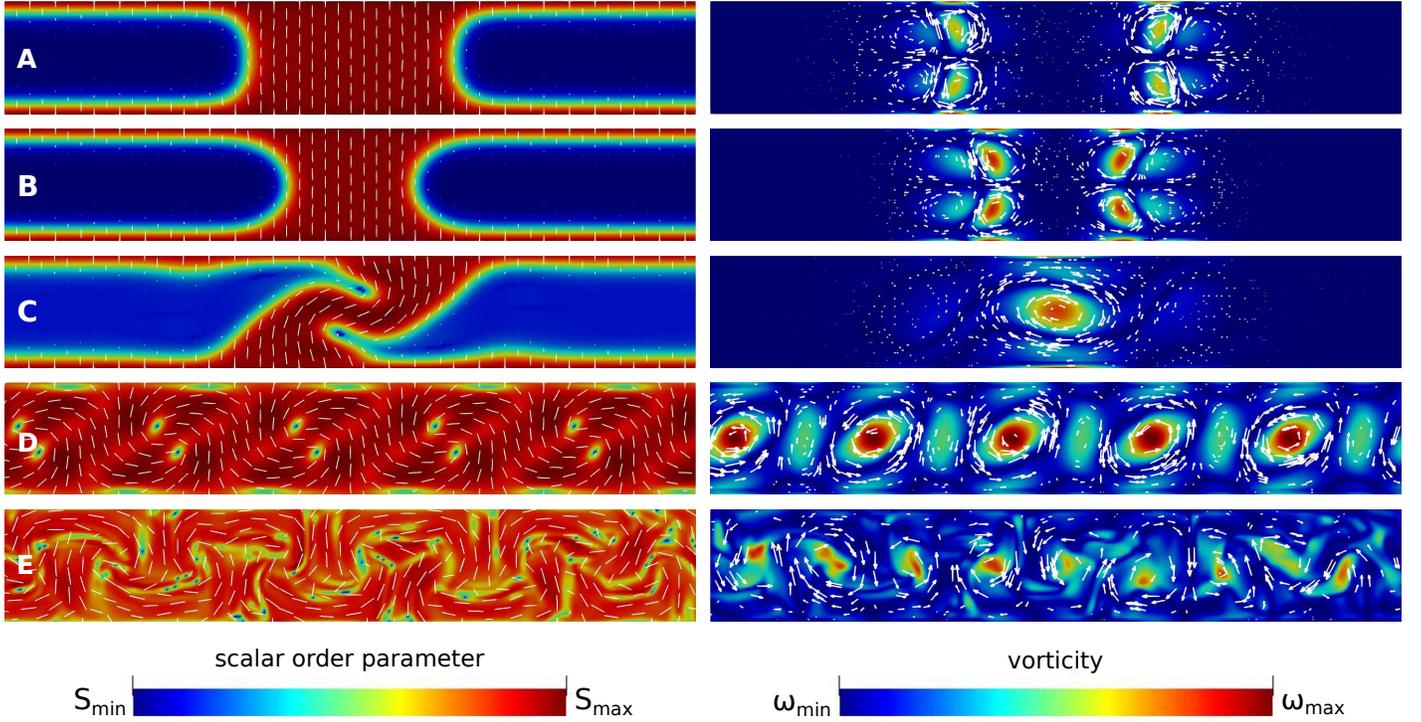}
\caption{Representative flow states in a channel with homeotropic anchoring at the walls at different activities. The system is at the coexistence temperature of the passive system in the channel ($\gamma_{NI}^{ch}$). The active nematic is flow aligning ($\xi=0.7$) and the screenshots were taken after $t=2\times 10^{6}$ iterations. A) For a contractile system with $\zeta=-0.002$, the nematic phase expands. B) For an extensile system with $\zeta = 0.0005$, the nematic phase contracts. C) At $\zeta=0.002$, the interfacial dancing state with one pair of defects is observed. D) At $\zeta=0.0075$, the system is in the dancing flow state (no interface). E) At $\zeta=0.05$, active turbulence is observed. Left: the lines represent the director field and the colors represent the scalar order parameter normalized to the minimum and maximum values, which are, from A to E, ($0$, $0.33$), ($0$, $0.33$), ($-0.03$, $0.34$), ($-0.10$, $0.45$) and ($-0.23$, $0.71$). Right: the lines indicate the direction of the velocity field and the norm of the vorticity is color coded, where the minimum and maximum values are, from A to E, ($0$, $1.86\times10^{-4}$), ($0$, $5.48\times 10^{-5}$), ($0$, $1.98\times 10^{-3}$), ($4.68\times 10^{-6}$, $0.0118$), ($1.53\times 10^{-6}$, $0.159$).}
\label{zeta-normal}
\end{figure*}

\section{Methods}
\label{methods-sec}

\subsection{Equations of motion}

We employ the Landau-de-Gennes free energy, $\mathcal{F} = \int_V f\,d^3 r $, to describe the passive liquid crystal at equilibrium. The free energy density, $f$, is the sum of two terms: $f = f_{bulk} + f_{el}$, where $f_{bulk}$ is the bulk free energy and $f_{el}$ is the elastic energy, i.e., the energy cost associated to distortions with respect to the uniform alignment of the director field. These energy densities are given by:
\begin{align}
    f_{bulk} =& \frac{A_0}{2}\left( 1- \frac{\gamma}{3} \right) Q_{\alpha \beta} Q_{\alpha \beta} - \frac{A_0\gamma}{3} Q_{\alpha \beta} Q_{\beta \gamma} Q_{\gamma \alpha}  \nonumber \\ &+  \frac{A_0\gamma}{4} (Q_{\alpha \beta} Q_{\alpha \beta} )^2, \nonumber \\ 
  f_{el} =& \frac{K}{2} (\partial _\gamma Q_{\alpha \beta}) (\partial _\gamma Q_{\alpha \beta}).
\end{align}
The tensor order parameter is assumed to be uniaxial, both in the initial conditions and in the interpretation of the results: $Q_{\alpha \beta} = S(n_\alpha n_\beta - \delta_{\alpha \beta}/3)$, where $S$ is the scalar order parameter and $n_\alpha$ is the $\alpha$-component of the director. The parameter $\gamma$ controls the magnitude of the order and depends on the temperature for thermotropic or on the density for lyotropic LCs, $K$ is the elastic constant and $A_0$ is a constant. In this one-constant approximation the anchoring at the NI interface is planar~\cite{p1995physics}. 
Here the Greek indices stand for Cartesian coordinates with sums over the terms with repeated indices. The coexistence between the bulk nematic and isotropic phases occurs at $\gamma_{NI}=2.7$ and $S=1/3$ as obtained by minimizing $f_{bulk}$. For $\gamma>\gamma_{NI}$ the bulk equilibrium phase is nematic while for $\gamma<\gamma_{NI}$ the bulk equilibrium phase is isotropic. 

The tensor order parameter evolves according to the Beris-Edwards equation:
\begin{align}
  \partial _t Q_{\alpha \beta} + u _\gamma \partial _\gamma Q_{\alpha \beta} - S_{\alpha \beta}(W_{\alpha\beta}, Q_{\alpha\beta}) = \Gamma H_{\alpha\beta} , 
  \label{beris-edwards}
\end{align}
where $\Gamma$ is the collective rotational diffusive constant. The co-rotational term, $S_{\alpha \beta}$, describes the response of the LC to gradients in the velocity field $\mathbf{u}$ and it is given by:   
\begin{align}
 S_{\alpha \beta} =& ( \xi D_{\alpha \gamma} + W_{\alpha \gamma})\left(Q_{\beta\gamma} + \frac{\delta_{\beta\gamma}}{3} \right) 
 + \left( Q_{\alpha\gamma}+\frac{\delta_{\alpha\gamma}}{3} \right)(\xi D_{\gamma\beta}-W_{\gamma\beta}) \nonumber \\&  - 2\xi\left( Q_{\alpha\beta}+\frac{\delta_{\alpha\beta}}{3}  \right)(Q_{\gamma\epsilon} \partial _\gamma u_\epsilon),
\end{align}
where $W_{\alpha\beta} = (\partial _\beta u_\alpha - \partial _\alpha u_\beta )/2$, $D_{\alpha\beta} = (\partial _\beta u_\alpha + \partial _\alpha u_\beta )/2$. The aligning parameter $\xi$ depends on the particles shape: it is positive for rod-like particles and negative for disk-like particles. The molecular field stands for:
\begin{align}
 H_{\alpha\beta} = -\frac{\delta \mathcal{F}}{\delta Q_{\alpha\beta}} + \frac{\delta_{\alpha\beta}}{3} \Tr \left( \frac{\delta \mathcal{F}}{\delta Q_{\gamma \epsilon}} \right).
\end{align}

The Navier-Stokes and continuity equations give the time evolution of the velocity field in terms of the stress tensor $\Pi_{\alpha\beta}$, which depends on the tensor order parameter:
\begin{align}
 \partial _t \rho + \partial_\alpha (\rho u_\alpha) =0,
\end{align}
\begin{align}
\rho \partial _t u_\alpha + \rho u_\beta \partial_\beta u_{\alpha} = \partial _\beta \Pi_{\alpha\beta}  + \eta \partial_\beta\left( \partial_\alpha u_\beta + \partial_\beta u_\alpha \right).
\label{navier-stokes}
\end{align}
The stress tensor is the sum of an active and a passive terms: $\Pi_{\alpha\beta} = \Pi^{\text{passive}}_{\alpha\beta} + \Pi^{\text{active}}_{\alpha\beta}$. In the simplest description of the hydrodynamics of active LCs, which accounts for the lowest 
order contribution with the appropriate symmetry, the active term is given by~\cite{PhysRevLett.92.118101}:
\begin{align}
 \Pi^{\text{active}}_{\alpha\beta} = -\zeta Q_{\alpha\beta},
 \label{active-pressure-eq}
\end{align}
while the passive one is:
\begin{align} 
 \Pi^{\text{passive}}_{\alpha\beta} =& -P_0 \delta_{\alpha\beta} + 2\xi \left( Q_{\alpha\beta} +\frac{\delta_{\alpha\beta}}{3} \right)Q_{\gamma\epsilon}H_{\gamma\epsilon} \nonumber \\ &- \xi H_{\alpha\gamma} \left( Q_{\gamma\beta}+\frac{\delta_{\gamma\beta}}{3} \right) - \xi \left( Q_{\alpha\gamma} +\frac{\delta_{\alpha\gamma}}{3} \right) H_{\gamma \beta} \nonumber \\ &- \partial _\alpha Q_{\gamma\nu} \,\frac{\delta \mathcal{F}}{\delta (\partial_\beta Q_{\gamma\nu})} + Q_{\alpha\gamma}H_{\gamma\beta} - H_{\alpha\gamma}Q_{\gamma\beta} .
 \label{passive-pressure-eq}
\end{align}
In these equations $\eta$ is the shear viscosity, $P_0$ is the hydrostatic pressure, and $\zeta$ is the activity parameter. Figs.\ref{active-particle-fig} A and B are a schematic representation of the velocity field for positive and negative $\zeta$. For $\zeta>0$, the velocity field resembles the one for pushers and the system is extensile, while for $\zeta<0$, it resembles the flow for pullers and the system is contractile.

\begin{figure}[t]
\center
\includegraphics[width=\linewidth]{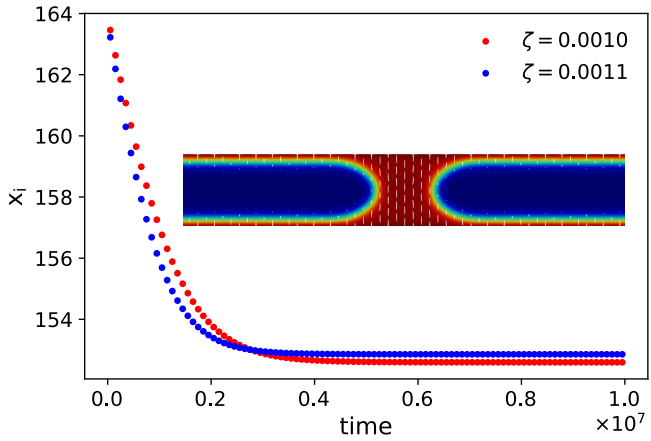}
\caption{Position of the interfaces of extensile nematics with different activities, at $\gamma_{NI}^{ch}$. The interfaces become static due to their mutual interaction, through the velocity field, when they are sufficiently close. The inset shows the scalar order parameter and the directors for the active nematic with $\zeta=0.001$ at $t=10^{7}$. The nematic is flow aligning with $\xi=0.7$. }
\label{stable-fig}
\end{figure}
\begin{figure}[t]
\center
\includegraphics[width=\linewidth]{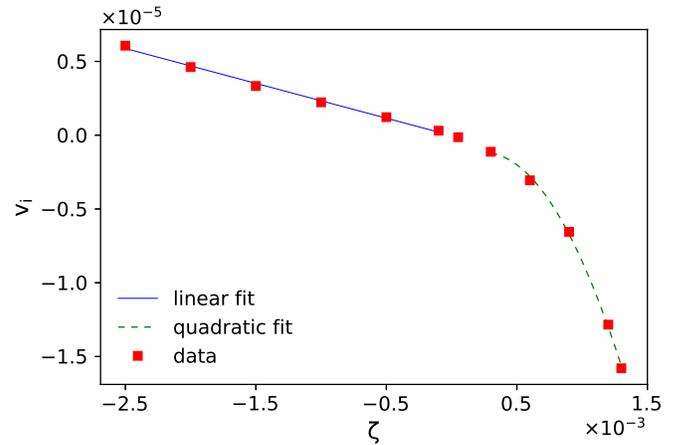}
\caption{Velocity of stable interfaces for flow aligning nematics, $\xi=0.7$, with different activities. The velocities are calculated when the interfaces are far apart, $t=50000$ to $t=150000$. The blue line is a linear fit that illustrates the linear regime of the interfacial velocity for contractile nematics. The dashed line is a quadratic fit showing that the linear regime for extensile nematics is greatly reduced.}
\label{ux-zeta-fig}
\end{figure}
\begin{figure*}[htb]
\center
\includegraphics[width=\linewidth]{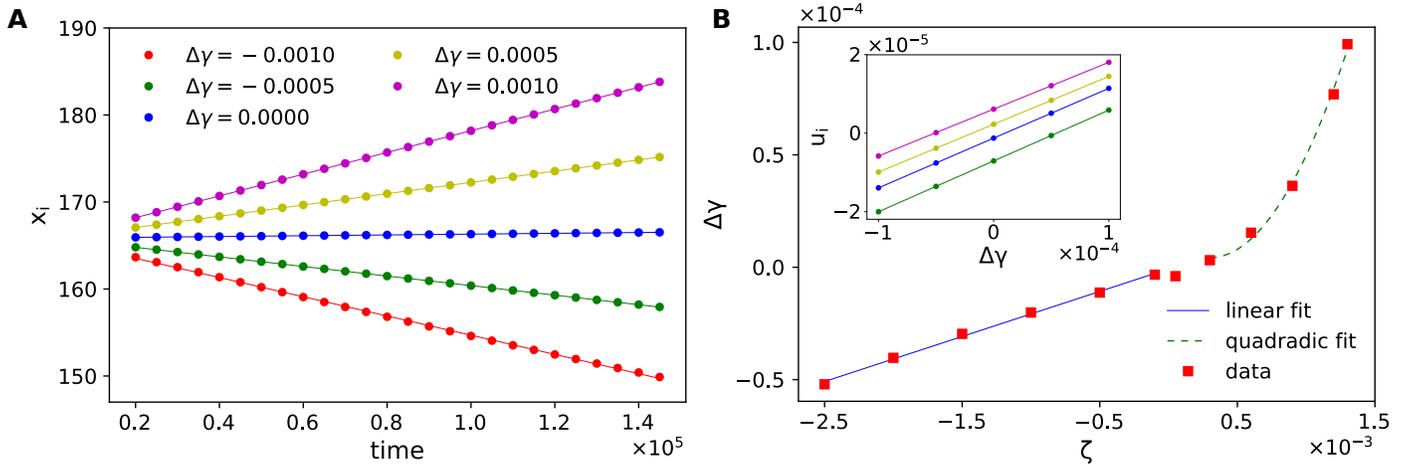}
\caption{Phase diagram of active nematics in a channel. A) Time evolution of the interface position at $y=L_Y/2$ at different temperatures for a contractile nemtic with $\zeta=-0.002$. The solid lines are linear fits from which the velocity is obtained. B) Phase diagram. $\Delta \gamma = \gamma - \gamma_{NI}^{ch}$ is the shift from the passive nematic isotropic temperature in the channel required to observe a static active interface for active nematics. The solid line is a fit illustrating a linear phase diagram for contractile nematics while the dashed line is a quadratic fit illustrating a more complex behaviour for extensile nematics. Above the coexistence line the system is nematic and below it is isotropic. The inset shows the interfacial velocity as a function of $\Delta \gamma$ at four activities with the respective linear fit shown in the same color (from top to bottom): -0.0025, -0.0010, 0.0003, 0.0009. The nematics are flow aligning with $\xi=0.7$.}
\label{zeta-dg-neg}
\end{figure*}

\subsection{Hybrid method}
\label{hybrid-method-sec}

To perform the numerical simulations, we solved the hydrodynamic equations of liquid crystals using a hybrid method of LB and FD. This approach is common in the literature and has successfully reproduced analytical and experimental results~\cite{Doostmohammadi2018,PhysRevLett.110.048303, marenduzzo07steady}. The LB has many advantages if compared to other computational fluid dynamics methods, namely the way complex boundary conditions are treated, the performance in parallel architectures and the possibility to include many physical models~\cite{kruger2016lattice, succi2018lattice, PhysRevB.96.184307} which justifies its use to simulate liquid crystals. The first approaches were fully LB, meaning that both the Navier-Stokes and Beris-Edwards equations were solved using LB~\cite{Denniston2002}\footnote{In fact, the LB solves the Boltzmann equation and recovers the macroscopic equations of motion, e.g. the Navier Stokes equation, in the macroscopic limit as can be shown through the Chapman-Enskog expansion~\cite{kruger2016lattice}.}. Later, the Beris-Edwards equations started to be solved with FD since this is more efficient in memory usage and furthermore eliminates spurious terms arising from the coupling between these two equations of motion in the Chapman-Enskog expansion~\cite{marenduzzo07steady}. In the LB part (to solve the Navier-Stokes equation) of the models used in the literature, the stress tensor was implemented in the equilibrium distribution and the D3Q15 lattice was used to discretize the velocity space~\cite{Denniston2002, denniston2004lattice, marenduzzo07steady}. There are additional challenges associated with simulation of interfaces in multiphase LB models as the control of spurious velocities and the difficulty to set the physical values of both viscosity and surface tension~\cite{0953-8984-25-24-245103}. To overcome these problems, we introduced three main improvements. First, we adopted the D3Q19 lattice since it is more isotropic than the D3Q15 and can resolve with finer resolution the angular distribution of the velocity field. Second, we use the multi-relaxation-time (MRT) collision operator, which has superior stability and accuracy and allows us to independently choose the relaxation rates of the hydrodynamic moments\cite{doi:10.1098/rsta.2001.0955}. The MRT is known to reduce, by orders of magnitude, the spurious velocities in the pseudopotential LB models, which are a class of multiphase and multicomponent models that suffer badly from this problem~\cite{PhysRevE.82.046708, AMMAR201773, PhysRevE.73.047701}. Third, in our model, the stress tensor is implemented in the force term, which reduces the spurious velocities in the free energy models~\cite{10.1142/S0217979203017448}.

The discrete form of the Boltzmann equation with the MRT collision operator reads as follows:
\begin{align}
  &f_i(\mathbf{x}+\mathbf{c}_i \Delta t, t+\Delta t) - f_i(\mathbf{x}, t) \nonumber \\ &= \mathbf{M}^{-1}\mathbf{R}\mathbf{M} [ f_i(\mathbf{x}, t) - f_i^{eq}(\mathbf{x}, t) ]\Delta t + \mathcal{S}_i,
\end{align}
where the index $i=\{0, \ldots, 18\}$ runs over the velocity vectors. The transformation matrix $\mathbf{M}$, and the relaxation matrix $\mathbf{R}$ are given in the Electronic Supplementary Information. The source term, $\mathcal{S}_i$, is given by Guo's forcing scheme, the equilibrium moments of which are also given in the Electronic Supplementary Information (see also Ref.~\cite{kruger2016lattice} for a detailed description of the method). The equilibrium distribution is the Maxwell-Boltzmann distribution expanded up to second order in Hermite polynomials~\cite{COELHO2018144}:
 \begin{align}
  f^{eq}_i = \rho w_i \left[ 1+ \frac{\mathbf{c}_i\cdot\mathbf{u}}{c_s^2} + \frac{(\mathbf{c}_i\cdot\mathbf{u})^2}{2c_s^4} - \frac{\mathbf{u}^2}{2c_s^2}  \right],
 \end{align}
 where $\mathbf{c}_i$ and $w_i$ are the velocity vectors and the discrete weights of the lattice, and $c_s=1/\sqrt{3}$ is the speed of sound in the D3Q19 lattice. The density $\rho$ and the macroscopic velocity $\mathbf{u}$ are calculated from the distribution functions: 
 \begin{align}
  \rho = \sum _i f_i, \quad \rho \mathbf{u} = \sum _i \mathbf{c}_i f_i + \frac{\mathbf{F}_i \Delta t}{2} .
  \label{density-vel-eq}
 \end{align}
Notice that the velocity passed to the FD method is the one calculated with Eq.~\eqref{density-vel-eq}, which is corrected by the term $\mathbf{F}_i \Delta t/2$ to ensure second order accuracy. The force is calculated as $F_\alpha = \partial _\beta (\Pi_{\alpha\beta} + \rho {c_s}^2 \delta_{\alpha\beta})$, where $\Pi_{\alpha\beta}$ is given by Eqs.~\eqref{active-pressure-eq} and ~\eqref{passive-pressure-eq}. In the FD, Eq.~\eqref{beris-edwards} is discretized on the same grid as the LB and solved through the predictor-corrector algorithm using second order differences. Therefore, both methods (FD and LB) are second order accurate and should provide consistent solutions of the equations of motion.

\section{Stable interface}
\label{stable-int-ec}

In this section, we simulate active stable interfaces (low activities) in a channel and compare the results with those of passive interfaces. We show that the active interfaces move with a velocity that depends on the activity and that they may be stabilized 
by a shift in the temperature.

\subsection{Channel setup and initial conditions}
\label{channel-setup-sec}

We start by simulating an open channel, with the nematic, between $x_1=3L_X/8$ and $x_2=5L_X/8$, and the  isotropic fluid elsewhere. The fluid is confined between two flat walls, at a fixed distance $L_Y$, with infinite homeotropic anchoring (see Fig.~\ref{setup-fig}) and periodic boundary conditions in the $x$ direction. The initial velocity is set to zero and the density to $\rho=1$ everywhere. The following parameters are used: $\tau=1.5$, $L_x\times L_Y\times L_Z= 270\times 45 \times 1$, $K=0.04$, $A_0=0.1$ and $\Gamma=0.34$. The other parameters change for each simulation and will be given in the caption of the corresponding figures. Our results are given in lattice units: the distance between nodes is $\Delta x = 1$ and the time step is $\Delta t = 1$ (see Refs.~\cite{Thampi_2015, Doostmohammadi2018} to transform these to physical units).

Since the interfacial profile follows approximately a hyperbolic tangent we initialize the scalar order parameter as:
\begin{align}
S = \frac{S_n}{2}\left[\tanh\left( \frac{x-x_1}{2 \lambda} \right) - \tanh\left(\frac{x-x_2}{2 \lambda}\right) \right],
 \label{S-initial-eq}
\end{align}
where $\lambda=\sqrt{27 K/A_0 \gamma}$ is the correlation length (equal to 2 at the bulk nematic-isotropic coexistence $\gamma=2.7$) and the scalar order parameter of the nematic phase, $S_n$, is computed through the minimization of the free energy:  
\begin{align}
S_n = \frac{\gamma + \sqrt{3(3\gamma^2-8\gamma )}}{4 \gamma}.
\label{Sn-eq}
\end{align}
$S_n =1/3$ at bulk coexistence $\gamma=2.7$. This expression for $S_n$ is used in the initial conditions and its value is fixed at the walls. We find the position of the interface, $x_{i}$, by fitting 
\begin{align}
 S = \frac{S_n}{2}\left[  1-\tanh\left(  \frac{x-x_{i}}{2 \lambda}   \right )   \right].
 \label{x-interface-eq}
\end{align}
to the interface on the right (the interface on the left is a mirror image in the stable regime). Other fields proportional to $S$, such as $Q_{yy}$, can also be used to obtain the position of the interface. In the passive case, the interface relaxes and becomes concave~\cite{refId0}, indicating that the nematic wets the channel walls. Indeed, a thin film of nematic is observed between the channel walls and the isotropic phase, for all the parameters considered here. In the simulations of active nematics, we turn the activity after 20000 iterations, allowing for the relaxation of the passive interface. 

\subsection{Passive interface in a channel}
\label{passive-sec}

In a liquid confined between two walls, the coexistence liquid-vapour temperature changes due to the interaction of the fluid with the walls, an effect known as capillary condensation~\cite{doi2013soft}. This also happens in nematics and has been called capillary nematisation. At the bulk NI transition temperature ($\gamma_{NI}$) the interface between the confined nematic and isotropic phases will move, leading to the expansion of the nematic if the walls favor the ordered phase~\cite{croxton1986fluid}. We calculate numerically the NI coexistence temperature under confinement $\gamma^{ch}_{NI}$, as the temperature where the passive interface is static. We start by evaluating the position of the interface at different times through a fit of $Q_{yy}$ using Eq.~\eqref{x-interface-eq}. Fig.\ref{passive-fig} shows the interfacial velocity at five different temperatures. Since the dependence is linear it is straightforward to calculate the NI coexistence temperature in the channel $\gamma^{ch}_{NI}\approx 2.68964$ as the temperature where the interfacial velocity vanishes. At temperatures $\gamma > \gamma^{ch}_{NI}$ the velocity is positive for the interface on the right (nematic expansion) and for $\gamma < \gamma^{ch}_{NI}$ the velocity is negative (nematic contraction). The interface is stable for a range of temperatures $\vert \Delta \gamma \vert < 0.02$ around $\gamma^{ch}_{NI}$. At temperatures higher or lower than these, one of the two phases becomes mechanically unstable, the interface disappears and the system becomes nematic ($\gamma > \gamma^{ch}_{NI}$) or isotropic ($\gamma < \gamma^{ch}_{NI}$). 
For the narrow channels considered here the strong anchoring at the walls dominates the interfacial anchoring and the anchoring at the interface is dictated by the wall anchoring (see Fig.\ref{active-particle-fig} C). 
\begin{figure}[t]
\center
\includegraphics[width=\linewidth]{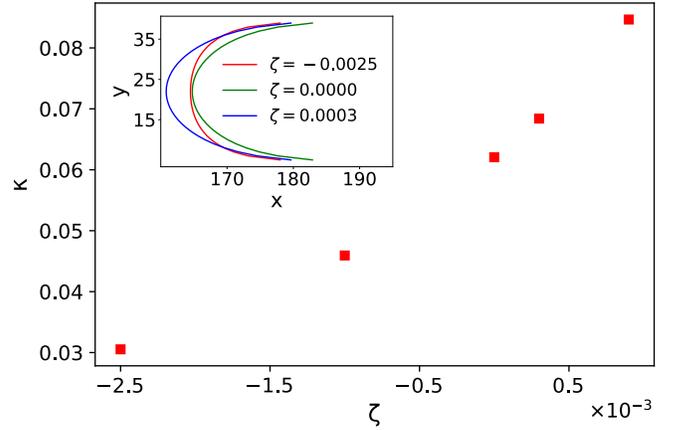}
\caption{Curvature at the center of static active interfaces stabilized by a temperature shift as a function of the activity, for flow aligning nematics with $\xi=0.7$. The inset shows the shape of the interface at three different activities. The curvature was calculated at $t=10^{7}$.}
\label{curvature-fig}
\end{figure}

\subsection{Active interface and stabilization by temperature}
\label{stabilization-temp-sec}

In order to proceed we consider the interface of an active nematic at the coexistence temperature of the passive system, $\gamma = \gamma^{ch}_{NI}$.
\begin{figure*}[t]
\center
\includegraphics[width=\linewidth]{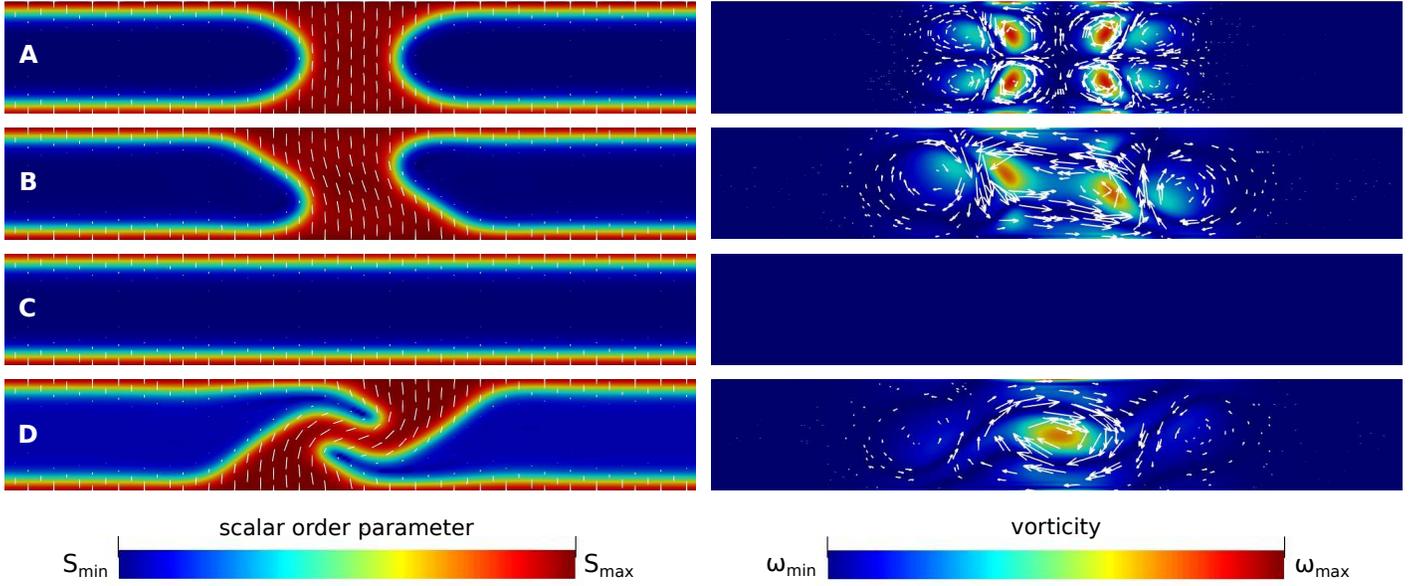}
\caption{Interfacial instabilities in extensile nematics with $\xi=0.7$. Screenshots were taken at $t=10^{7}$ for systems with different activities at $\gamma = \gamma^{ch}_{NI}$. A) At $\zeta=0.0011$ the interface is stable and static due to the interaction between the two interfaces. B) At $\zeta=0.0012$, the interface breaks the symmetry with respect to the centre of the channel but remains static. C) At $\zeta=0.0013$, an interfacial instability is formed and the system becomes disordered (the interface disappears). D) At $\zeta=0.0014$ the interface is highly asymmetric but still static. Finally, at $\zeta=0.0015$ an interfacial dancing state is observed. Left: the lines represent the director field and the colors represent the scalar order parameter normalized to the minimum and maximum values, which are, from A to D, ($-2.73\times 10^{-3}$, $0.33$), ($-4.36\times 10^{-3}$, $0.33$), ($8.45 \times 10^{-5}$, $0.33$), ($-0.0291$, $0.33$). Right: the lines indicate the direction of the velocity field and the norm of the vorticity is color coded, where the minimum and maximum values are, from A to E, ($0$, $1.35\times 10^{-4}$), ($0$, $2.76\times 10^{-4}$), ($0$,$0$), ($0$, $1.11\times 10^{-3}$).}
\label{instability-fig}
\end{figure*}
\begin{figure}[t]
\center
\includegraphics[width=\linewidth]{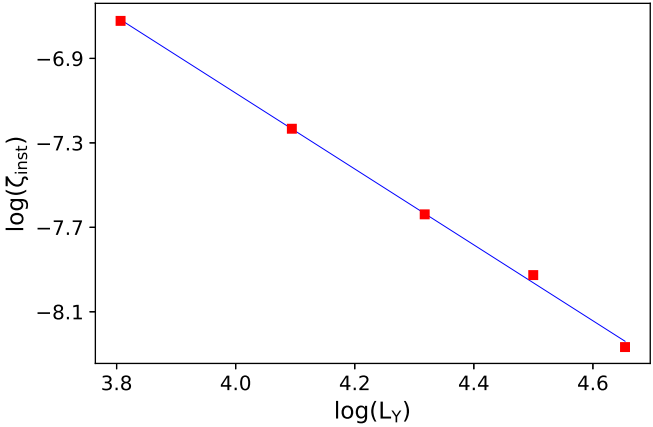}
\caption{Activity of the interfacial instability as a function of the channel height for flow aligning nematics with  $\xi=0.7$. The slope of the linear fit is -1.8.}
\label{size-fig}
\end{figure}
\begin{figure*}[t]
\center
\includegraphics[width=\linewidth]{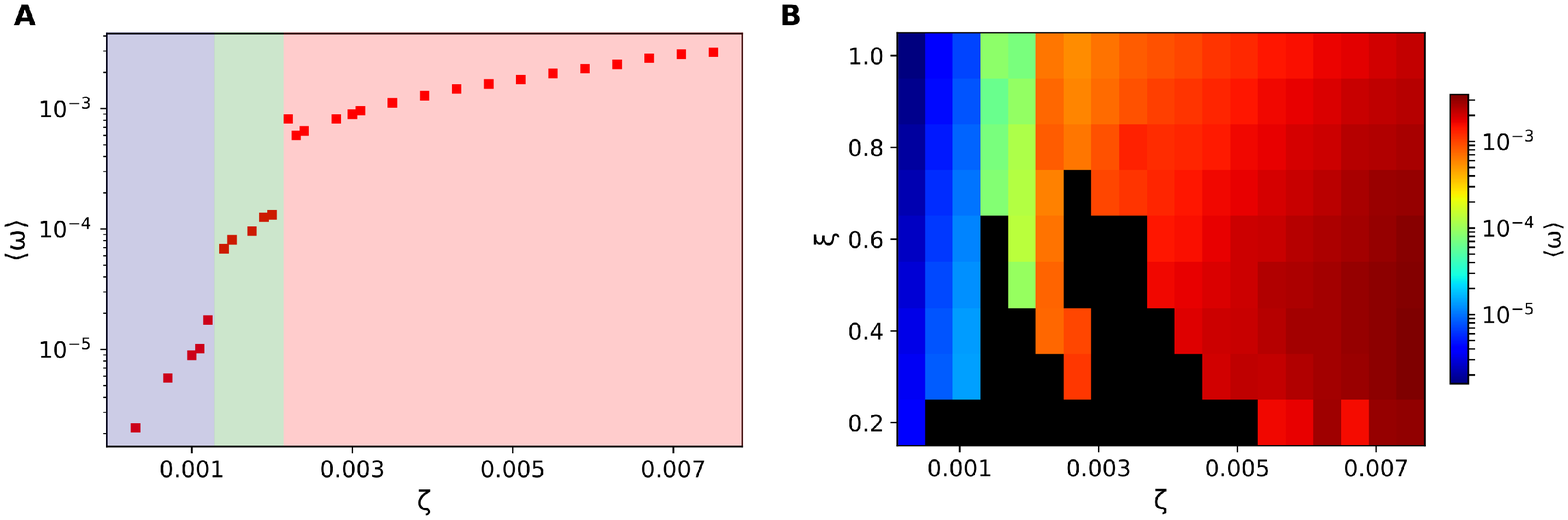}
\caption{Vorticity of systems with different activities at $t=2\times10^{6}$. A) Extensile nematics with $\xi =0.7$, at higher resolution (more points) than the color map; B) Color map for extensile nematics with different $\xi$'s. The blue region corresponds to the stable interface moving with constant velocity; the green region corresponds to the interfacial dancing state; the red region corresponds to dancing and turbulent states (no interface); the black region corresponds to systems with zero vorticity (isotropic system and no interface). }
\label{xi-zeta-fig}
\end{figure*}

The orientation of the directors close to the interface is determined by a competition between the anchoring at the walls~\cite{batista2014effect} and the interfacial passive and active anchorings~\cite{C7SM00325K}. The latter tends to align the directors parallel to the interface in extensile systems and normal to it in contractile ones. In narrow channels with strong anchoring, as the ones considered here, the effect of the wall anchoring is dominant and the director field is uniform (Fig. \ref{active-particle-fig} C). In wider channels and/or for weak anchoring at the walls the directors are expected to align parallel to the interface, both in passive LCs with a single elastic constant $K$ as well as in extensile nematics with the same elasticity (Fig. \ref{active-particle-fig} D). 

At low activities a stable interface is formed (Fig.~\ref{zeta-normal} A and B). The interface moves with constant velocity and constant shape, while the bulk remains at rest. The domain of the nematic phase expands for contractile systems ($\zeta<0$) and contracts for extensile ones ($\zeta>0$). This behavior is counterintuitive since bulk extensile systems favor nematic ordering while contractile ones disfavor it. 

To understand this result, we calculate the active forces at the interface. Projecting the active force $\mathbf{F}^{\text{active}} = -\zeta \nabla \cdot \mathbf{Q}$ on the outward normal to the NI interface, $\mathbf{m}$, we have:
\begin{align}
\mathbf{F}^{\text{active}}_{\bot} &= -\zeta \Big[ (\mathbf{n}\cdot\mathbf{m}) (\mathbf{n}\cdot \boldsymbol{\nabla} S + S \, \boldsymbol{\nabla} \cdot \mathbf{n} ) \nonumber \\ & + S \,\mathbf{m}\cdot (\mathbf{n}\cdot \boldsymbol{\nabla})\mathbf{n} - \frac{1}{3} \textbf{m}\cdot \boldsymbol{\nabla} S \Big]  \textbf{m}.
\label{active-force-complete}
\end{align}
An estimate of the force is easily obtained by assuming that the interface is circular, with radius $R$, and that the director is 
parallel to it, $\mathbf{n}\cdot\mathbf{m}=0$ (Fig. \ref{active-particle-fig} E), which is reasonable in the center of the channel.
 The normal force is then given by the sum of two terms:
\begin{align}
 \mathbf{F}^{\text{active}}_{\bot} = -\zeta \left(   \frac{\vert\boldsymbol{\nabla}S \vert}{3} + \frac{S}{R} \right) \mathbf{m}.
 \label{active-force}
\end{align}
The first of these terms is the active force due to the gradient of the nematic order, while the second is due to the interfacial curvature. For extensile nematics, the normal active force acts inwards, while for contractile ones the force acts outwards, revealing the mechanism for the contraction and the expansion of the nematic phase observed in the simulations. In addition, the force is linear in $\zeta$ and it increases as $R$ decreases. We note that the second term of the active force, which increases the curvature of a curved interface with planar anchoring, acts in a similar way to the active forces that increase bend fluctuations in extensile nematics and are ultimately responsible for the bend instability of unconfined extensile nematics. 
By contrast, the first term of the force, proportional to $\vert\boldsymbol{\nabla}S \vert$, is absent deep in the nematic, where the scalar order parameter is nearly uniform. At the interface, however, this term becomes increasingly important as the curvature decreases, and it is the only active force at a flat interface. Similar active forces were reported in Ref.~\cite{PhysRevLett.113.248303} for interfaces in active nematic emulsions, characterized by an additional conserved order parameter. One important difference between our results and those of Ref.~\cite{PhysRevLett.113.248303} is that the interfaces considered here are stable due to the confinement in the channel.  

In what follows we show that the active interfacial motion can be stopped by a shift in temperature, in a way reminiscent of the behaviour of passive nematic-isotropic interfaces near coexistence.  

We start by noting that, for extensile nematics $\zeta>0$ the interfaces move towards each other and, when they are sufficiently close, the velocity field of one pushes the other rendering the interfaces static without the need for a temperature shift (see Fig.~\ref{stable-fig}). This effect does not occur for contractile systems $\zeta<0$ as the two interfaces move in opposite directions. We are interested in the dynamics of a single interface, far from surfaces or other interfaces. Thus, in Fig.~\ref{ux-zeta-fig} we plot the interfacial velocity measured when the interfaces are far apart and the interaction between them is negligible.

For each system (or activity), we changed the temperature around $\gamma^{ch}_{NI}$ and measured the interfacial velocity. We found that the interfacial velocity is linear in $\Delta \gamma$, for both contractile and extensile nematics. It is then straightforward to calculate the shift $\Delta \gamma$ required to observe a static interface for a given activity (Fig.~\ref{zeta-dg-neg}A). In Fig.~\ref{zeta-dg-neg}B we plot the phase diagram for active nematics in a channel, as the temperature shift, $\Delta \gamma$, that renders the interface static. For contractile nematics the coexistence line is approximately linear. For extensile ones, the linear relation is observed in a narrow range of activities becoming quadratic afterwards. This quadratic behaviour correlates with the quadratic dependence of the interfacial velocity on the activity observed in extensile nematics (Fig.~\ref{ux-zeta-fig}).

In order to check that the combined effects of temperature and activity lead to a static interface, we performed longer simulations (up to $t=10^7$ time steps) at the values of $\Delta\gamma$ given in Fig.~\ref{zeta-dg-neg}B for different activities. We found that the interfacial velocities are indeed very close to zero at these temperatures (which could be refined iteratively). 

The shape of the static interface changes with the activity and is plotted in the inset of Fig.~\ref{curvature-fig}. Although there is no net flow under these conditions, the vortices near the interface (see Fig.~\ref{zeta-normal}) become stronger as the activity increases. We quantify the dependence of the static interfacial shape on the activity through the curvature at the center of the channel ($y=L_Y/2$). The interfacial curvature is defined as:
\begin{align}
 \kappa = \frac{\vert x_{i}^{\prime\prime}(y)     \vert }{\left[  1+(x_{i}^\prime (y))^2  \right]^{\frac{3}{2}}}.
\end{align}
and is plotted in Fig.~\ref{curvature-fig}, where $x_{i}(y)$ is the position of the interface. For extensile nematics the curvature increases with the activity. Assuming that the interface is circular, we can estimate its radius and find: $R (\zeta=-0.0025) \approx 33$ and $R (\zeta=0.0009) \approx 12$. 
 
We can now understand why the interfacial velocity has a linear regime and a non linear one, Fig. \ref{ux-zeta-fig}. When the curvature is small (large $R$), the second term in Eq. \eqref{active-force} is small and the force is approximately linear in $\zeta$. When the curvature increases (small $R$), the second term in Eq. \eqref{active-force}, which depends on the activity (Fig. \ref{curvature-fig}), is relevant and the active force becomes non-linear.

\section{Flow states and interfacial instability}
\label{flow-states-sec}

In Fig.~\ref{zeta-normal} we give an overview of the different states observed as the activity increases, from contractile to 
extensile nematics. The top pannels (A and B) correspond to contractile (A) and extensile (B) nematics where a stable interface, with fixed shape, was found to move at constant velocity in the channel. These states are observed at low activities. At higher activities, the interface of extensile nematics exhibits an instability. A complex interface with a persistent saptio-temporal structure (panel C) is observed for a range of activities beyond the instability. At even higher activities the interface cannot be sustained, as the isotropic phase in the channel becomes unstable, and complex flow states, dancing D and active turbulence E, are observed in line with earlier reports ~\cite{Doostmohammadi2018, 10.1039/C8SM02103A}. 

Stable interfaces have a shape that is invariant in time and mirror symmetric with respect to the centre of the channel. An interfacial instability occurs when the system breaks one or both of these symmetries (Fig.~\ref{instability-fig}) at a specific value of the activity, $\zeta_{inst}$. For the initial conditions described above, the interfaces move towards each other, interact through the velocity field and become static as illustrated in Fig.~\ref{stable-fig}. Fig.~\ref{instability-fig}B shows that at $\zeta_{inst} \approx 0.0012$ the interfaces break the mirror symmetry and one of the two vortices close to the interface becomes stronger. At $\zeta = 0.0013$, the system becomes isotropic and the interface disappears. This suggests that the system is multistable in this range
of parameters and that the final state may be affected by finite size effects. At $\zeta=0.0014$ (see Fig.~\ref{instability-fig} C) a strongly distorted interface reappears. The instability is driven by the active force that increases with the interfacial curvature which in turn increases the curvature of the interfaces with planar anchoring. This instability is closely related to the bend instability for extensile systems reported previously in unconfined active nematic emulsions~\cite{PhysRevLett.113.248303,C7SM00325K}. With one important difference: the active force that drives the interfacial instability is also responsible for the re-appearence of the nematic phase and the interfacial state persists at long times. 

A distinct instability in active nematics, where the fluid starts moving coherently across the channel, was reported deep in the ordered phase, at higher activities that scale with the inverse of the square of the channel height~\cite{VoituriezEPL2005}. This scaling was verified numerically in the non-linear regime in Ref.~\cite{marenduzzo07steady} and is also approximately verified for the interfacial instability. In Fig.~\ref{size-fig} we plot the activity at threshold $\zeta_{inst}$ for channels with different heights. For each channel $L_Y$, we calculated $\gamma^{ch}_{NI}$ as the passive NI temperature depends on the channel height. We found that the value of the threshold activity decreases approximately with ${L_Y}^2$ in line with the scaling reported previously. This is to be expected as the active length associated with the active vortices scales with the ratio of the elastic to the active forces, $\sqrt{K/\zeta}$ and the activity number, which is the ratio of the channel height to the active length, is nearly constant, $A=L_Y \sqrt{\zeta/K} \approx 7.8$, at the interfacial instability threshold.  

At higher activities, in the slab geometry, an interfacial dancing state (Fig.~\ref{zeta-normal}C) is observed. This state is characterized by two NI interfaces and a pair of $+1/2$ defects moving in complex oscillatory trajectories. The defects leave one interface into the nematic and disappear from it at the other interface. This state resembles the bulk dancing state, confined by the NI interfaces. The interfaces are no longer concave with constant curvature, as the stable interfaces discussed in Sec.~\ref{stable-int-ec}, but they are not transient. The interfacial dancing state is stable over a finite range of activities: for a flow aligning nematic with $\xi=0.7$ and a channel with $L_Y=45$, the state is observed for activities in the range $0.0015 < \zeta < 0.0021$.

At even higher activities, the interface disappears (the isotropic phase becomes unstable) and the previously reported dancing (Fig.~\ref{zeta-normal}D) and active turbulent states (Fig.~\ref{zeta-normal}E) are observed~\cite{Doostmohammadi2018, 10.1039/C8SM02103A}. 

We used the spatial average of the vorticity norm in the steady state, calculated at $t=2\times10^{6}$, as an order parameter to characterize the states of extensile nematics: $\langle \omega  \rangle$, where $\boldsymbol{\omega} = \nabla \times \mathbf{u}$. In Fig.~\ref{xi-zeta-fig}A, we plot the average vorticity as a function of the activity for nematics with $\xi=0.7$. 
The results reveal that the average vorticity $\langle  \vert \mathbf{\omega} \vert \rangle$ increases as the activity increases. More interestingly the vorticity exhibits jumps between dynamical states. The first jump, at $\zeta\approx 0.0012$, occurs between the stable interface and the interfacial dancing states. As discussed previously, we have observed multi-stability and a sequence of intermediate states (see Fig.~\ref{instability-fig}) including isotropic states with zero vorticity, in the transition region. States with zero vorticity are represented by the black squares in the right panel of the figure, Fig.~\ref{xi-zeta-fig}B. The second jump occurs between the interfacial dancing and the dancing states where the nematic fills the channel and the interface is no longer stable. In Fig.~\ref{xi-zeta-fig}B we plot the vorticity at $t=2\times 10^{6}$ for active nematics with different alignment parameters and activities. At the bulk NI temperature the flow is tumbling if the alignment parameter $\xi <0.43$~\cite{marenduzzo07steady} but the stability of the flow states does not depend strongly on it. However, most of the black squares (isotropic) are in the flow tumbling regime.

\section{Other channels: planar anchoring and closed channel} 
\label{other-setups-sec}

\begin{figure}[t]
\center
\includegraphics[width=\linewidth]{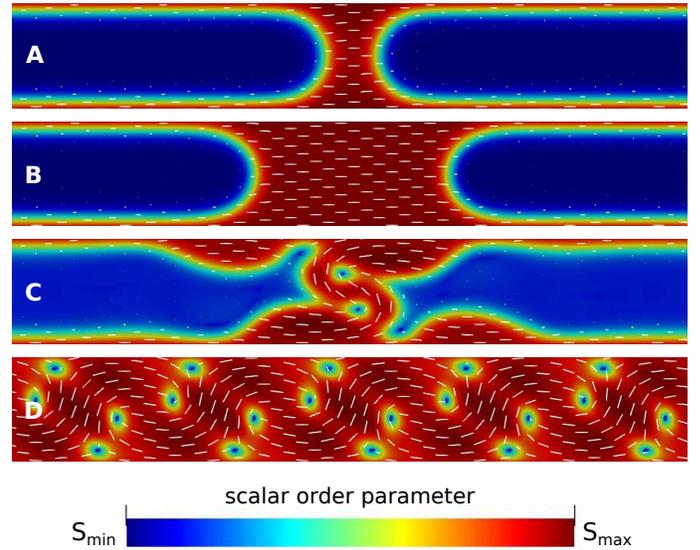}
\caption{Channel with planar anchoring. A) Contraction of the nematic phase for contractile systems with $\zeta=-0.001$ after 250000 iterations (the final state is isotropic); B) Expansion of the nematic phase for extensile systems $\zeta = 0.001$ after 250000 iterations (the final state is nematic); C) Interfacial dancing state $\zeta=0.0025$ at $t=5\times10^6$; D) Dancing state $\zeta=0.003$ at $t=5\times10^6$. Here $\xi=0.5$. The lines represent the director field and the colors represent the scalar order parameter normalized to the minimum and maximum values, which are, from A to D, ($-6.64\times 10^{-5}$, $0.33$), ($-2.17\times 10^{-3}$, $0.33$), ($-0.0289$, $0.34$), ($-0.0637$, $0.37$).}
\label{parallel-fig}
\end{figure}
\begin{figure}[t]
\center
\includegraphics[width=\linewidth]{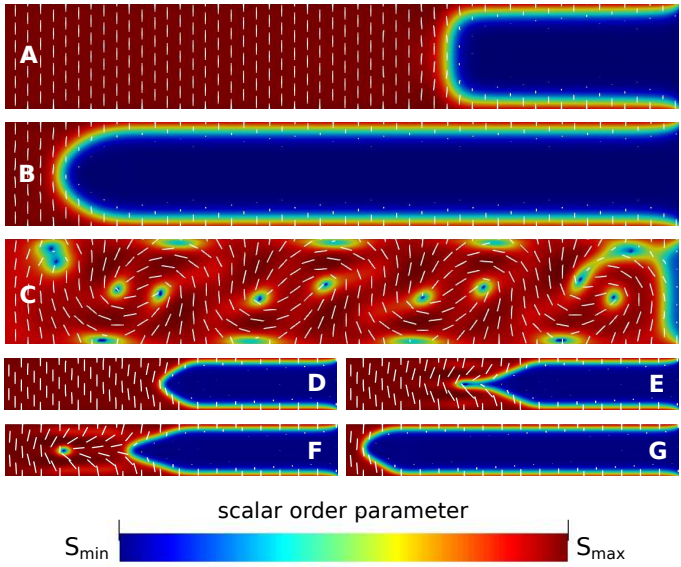}
\caption{Closed channel with homeotropic anchoring. The figures from A to C illustrate the flow states in a closed channel at $t=5\times10^6$. A) Expansion of the nematic phase for contractile systems $\zeta=-0.005$; B) Contraction of the nematic phase for extensile systems with a symmetric but static interface close to the left border $\zeta = 0.001$; C) Dancing state at $\zeta=0.003$. The panels D to G depict the interfacial instability in the closed channel for an extensile system with  $\zeta=0.002$ at: D) $t=15000$, E) $t=30000$, F) $t=40000$, G) $t=5\times 10^6$. Note that the instability in the closed channel occurs by breaking time invariance through the creation of a defect that relaxes the interfacial curvature. The interface breals the mirror symmetry only when it interacts with the left wall. Here $\xi=0.5$. The lines represent the director field and the colors represent the scalar order parameter normalized to the minimum and maximum values, which are, from A to D (colors for D-G are the same), ($-8.07\times 10^{-4}$, $0.33$), ($-1.96\times 10^{-3}$, $0.33$), ($-0.0719$, $0.38$), ($-0.0564$, $0.34$). }
\label{closed-fig}
\end{figure}

Finally, we consider other channels to investigate the effects of the anchoring at the walls and the open boundary condition. 

We start by considering planar channels with strong anchoring and scalar order parameter given by Eq.~\ref{Sn-eq}. For narrow channels the confined nematic is uniform and the alignment by the walls dominates the interfacial and the active anchorings. We 
initialize the director field in this configuration. We consider a slab geometry with the nematic in the center of the channel surrounded by isotropic. All other parameters are as in Sec.~\ref{channel-setup-sec}. Fig.~\ref{parallel-fig}, illustrates a sequence of states that resemble those found for homeotropic channels. 

At low activities, the interface is stable and propagates with constant velocity depending on the activity (see Fig.~\ref{parallel-fig} A and B). However, the nematic phase contracts for contractile systems and expands for extensile ones.
As before we use Eq. \eqref{active-force-complete}, to calculate the active force for a circular interface with a normal director field, $\mathbf{n}\cdot \mathbf{m}=1$ (Fig. \ref{active-particle-fig} F). The active force is now:
\begin{align}
 \mathbf{F}^{\text{active}}_{\bot} = \zeta \left(   \frac{2\vert\boldsymbol{\nabla}S \vert}{3} + \frac{S}{R} \right) \mathbf{m}.
 \label{active-force-planar}
\end{align}
Clearly, the active forces at interfaces with planar and homeotropic anchoring point in opposite directions (see Eq. \eqref{active-force}). In planar channels, the active force at the interface is also larger, in line with the larger velocities observed in the simulations.

In Fig.~\ref{parallel-fig} C, we illustrate the interfacial dancing state observed in the planar channel. We note the formation of islands of nematic close to the boundaries and the presence of $+1/2$ defects that move in directions opposite to those observed in the homeotropic channel (as a result of the opposite signs of the vorticities). The dancing state, shown in Fig.~\ref{parallel-fig} D, is similar in both channels. 

Finally, we consider a closed channel with only one interface (see Fig.~\ref{closed-fig}). The left boundary is nematic with 
planar anchoring and the right one is isotropic, with no slip boundary conditions at the four walls. The system is initialized with 
the left half of the channel in the uniform nematic state with directors parallel to the interface and the right half isotropic. We note that the stable interface behaves in the same way as described in Sec.~\ref{stable-int-ec}. Furthermore, the interfacial dancing state appears even when the channel is closed (Fig.~\ref{closed-fig} C) but manifests itself through a different route (Fig.~\ref{closed-fig} D to G): The interface does not break the mirror symmetry but becomes increasingly curved until a defect is formed and is ejected into the nematic phase. When the interface approaches the left rigid boundary it finally breaks the mirror symmetry and  becomes asymmetric as in Fig.~\ref{instability-fig}.

\section{Summary and conclusion}\label{conclusions}

To summarize, we studied the behavior of nematic-isotropic interfaces in active liquid crystals confined in a narrow channel. We
found that, at high activities, the interface disappears and the system becomes nematic in the previously reported dancing or 
active turbulent flow states. 
At lower activities, there are two flow states where an interface is present: the stable interface and the interfacial dancing state. In the stable interface regime with homeotropic anchoring, the right interface propagates forwards for contractile nematics and backwards for extensile ones, as a result of the active force at the interface.

The moving active interfaces may be stopped by a shift in the temperature. We calculated this shift for a wide range of activities and checked the robusteness of this stabilization mechanism by performing longer simulations. The static active interfaces are flatter than the passive ones for contractile nematics and more strongly curved for extensile ones. The curvature increases with the activity and, above a critical curvature, the interface becomes unstable. In addition, we observed an interfacial dancing state similar to the unconfined dancing state. In this state, +1/2 defects are continuously formed at the IN interface and are ejected into the bulk nematic. We also explored other channels and initial conditions to verify the robustness of the results.
  
To perform the simulations, we used an improved hybrid model based on the lattice Boltzmann method and on a predictor-corrector FD scheme. Among the improvements are: the use of a more isotropic lattice (D3Q19); a multi-relaxation-time collision operator, which is more accurate and stable since it allows the choice of different relaxation rates for the hydrodynamic moments; implementation of the stress tensor in the force term, eliminating spurious velocities. Those improvements were required to obtain a reliable description of the interfacial dynamics since these systems are prone to spurious numerical effects. 

A final word on experiments. Although the motivation of the work reported here was theoretical, the relevance of interfaces in active nematic experiments is clear. As mentioned in the introduction the active turbulent state may be observed at temperatures above the passive NI transition rather than deep in the bulk nematic phase. Under these conditions, the role of  interfaces may affect transient or steady dynamical states, as described here. Indeed, recently an experimental study of the propagation of active-passive interfaces in bacterial swarms has been reported~\cite{Patteson2018}. The methods and models proposed in our work may be used to address a number of questions raised by these and related experiments.      

\section*{Conflicts of interest}
There are no conflicts to declare.

\section*{Acknowledgements}

We acknowledge financial support from the Portuguese Foundation for Science and Technology (FCT) under the contracts: PTDC/FIS-MAC/28146/2017 (LISBOA-01-0145-FEDER-028146) and UID/FIS/00618/2019. Margarida Telo da Gama (MTG)
would like to thank the Isaac Newton Institute for Mathematical Sciences for support and hospitality during the program ”The mathematical design of new materials” where most of this work was carried out. This program was supported by EP-SRC 
Grant Number: EP/R014604/1. MTG participation in the program was supported in part by a Simons Foundation Fellowship. We thank Prof. Mykola Tasinkevych for the fruitful discussions.



\balance


\bibliography{rsc} 
\bibliographystyle{rsc} 

\end{document}